\documentclass{svproc}
\usepackage{graphicx}%
\usepackage{multirow}%
\usepackage{amsmath,amssymb,amsfonts}%
\usepackage{mathrsfs}%
\usepackage{xcolor}%
\usepackage{textcomp}%
\usepackage{manyfoot}%
\usepackage{booktabs}%
\usepackage{algorithm}%
\usepackage{algorithmicx}%
\usepackage{algpseudocode}%
\usepackage{listings}%
\usepackage{tikz}
\usetikzlibrary{arrows.meta}
\usepackage{url}

\def\orcidID#1{\unskip$^{[#1]}$} % added MR 2018-03-10

\usepackage{rotating}
\usepackage{array}
\usepackage{tabularx}

\usepackage{enumitem}
\setlist[itemize]{noitemsep, topsep=0pt}

\usepackage[normalem]{ulem}
\definecolor{revisionorange}{RGB}{225,120,0}
\definecolor{revisiongreen}{RGB}{0,135,55}
\definecolor{revisiongrey}{RGB}{135,135,135}

\begin{document}
\mainmatter              % start of a contribution
\title{Graphlet-based Decentralized Finance Transaction Method Classification}
%
%\titlerunning{Running title}  % abbreviated title (for running head)
%                                     also used for the TOC unless
%                                     \toctitle is used
%
\author{Célestin Coquidé\inst{1}\orcidID{0000-0001-8546-6587} \and Alessio Brini\inst{2}\orcidID{0000-0003-1121-818X} \and Natkamon Tovanich\inst{3}\orcidID{0000-0001-9680-9282}}
\authorrunning{Coquidé, Brini, and Tovanich} % abbreviated author list (for running head)

\institute{UTINAM Institute, UMR 6213 CNRS, Université Marie et Louis Pasteur, Besançon, France \email{coquide31@gmail.com}
\and Duke University, Durham, NC, USA \email{alessio.brini@duke.edu}
\and TU Wien, Vienna, Austria \email{natkamon.tovanich@tuwien.ac.at}}

\maketitle              % typeset the title of the contribution

\begin{abstract}
Decentralized Finance (DeFi) transactions involve complex interactions among accounts, smart contracts, and tokens. We investigate whether network topology can identify transaction methods that describe DeFi operations. We extract directed graphlet features to classify the 30 most frequent methods. Our results show that graphlet features complement event log counts in improving classification performance. Using SHapley Additive exPlanations (SHAP), we further derive model explanations that reveal method-specific graphlet structural patterns.
\end{abstract}
\section{Introduction}\label{sec:introduction}

Decentralized Finance (DeFi) provides financial services such as lending, borrowing, and trading without intermediaries through programmable smart contracts. Users interact with DeFi protocols through \textit{transactions} that invoke \textit{methods} implemented in smart contracts. Such methods may involve multiple tokens and contracts, while sequences of methods can be composed into more complex strategies~\cite{kitzler2023}. For instance, a transaction calling an aggregator contract may route trades through several decentralized exchange protocols to obtain a more favorable exchange rate.

Transaction methods provide a semantic description of the operations performed, but human-readable labels rely on public contract interfaces, which do not cover all transactions. Inferring transaction methods from observed execution behavior can help identify, inspect, and validate on-chain financial activities, particularly when explicit method information is unavailable or uncertain.

Network representations provide a natural framework for capturing these interactions, with accounts represented as nodes and token transfers as directed edges. In addition, smart contracts emit \textit{event logs} during execution, providing another observable description of transaction behavior. While event logs capture operations explicitly emitted by contracts during the execution, token transfer networks describe how assets move among participating entities.

In many real-world networks, small recurring substructures, commonly referred to as network motifs or graphlets, play important roles in representing the organization and functioning of complex systems~\cite{milo2002,przulj2007,sarajlic2016}. In this study, we investigate whether such structures can be used to characterize and identify DeFi transaction methods. Our hypothesis is that different transaction methods produce distinctive combinations of these structural building blocks.

Our contributions are threefold:
\begin{enumerate}
    \item We introduce a \textbf{topology-based representation of DeFi transactions} as directed, attributed multigraphs and extract graphlet-based features enriched with account types, token types, and categories.
    \item We evaluate \textbf{transaction-method classification} on the 30 most frequent methods, comparing graphlet-based features, event log count features, and their combination.
    \item We use \textbf{SHapley Additive exPlanations (SHAP)} to aggregate feature contributions at the graphlet level, revealing interpretable structural patterns associated with different DeFi methods.
\end{enumerate}

Our results show that graphlet topology captures a distinct dimension of DeFi transaction behavior that complements event log information. Our work contributes to the methodology and empirical evidence for characterizing DeFi operations, which is a first step toward understanding and decomposing complex transactions into interpretable structural building blocks of DeFi.

\section{Related Work}\label{sec:related_work}

Network motifs and graphlets characterize complex networks through local substructures. Motifs capture patterns occurring more frequently than expected in randomized networks~\cite{milo2002}, while graphlets represent small non-isomorphic subgraphs and have been introduced to compare biological networks~\cite{przulj2007} and extended to directed networks~\cite{sarajlic2016}.

In blockchain networks, such approaches have been used to study transaction and user behavior, including detecting Bitcoin mixing services and ransomware~\cite{akcora2020bitcoinheist,wu2022detecting}, characterizing user interactions~\cite{kosyfaki2019flow}, and detecting phishing scam activities~\cite{liu2024fishing,wang2023phishing}.
Structural and temporal motifs have also been used to analyze recurring behaviors in cryptocurrency ecosystems~\cite{aoyama2022cryptoasset,arnold2024insights}.
However, these studies mainly focus on analyzing user, service, ecosystem, or illegal behaviors rather than operations performed in DeFi transactions.

Machine learning has increasingly been applied to blockchain data~\cite{azad2025machine}, with applications ranging from entity identification~\cite{tovanich2023fingerprinting} to fraud detection~\cite{ma2025DefiFraud,rudrawar2025DeFiFraud}. However, these works focus on predictive performance rather than model explainability. Explainable artificial intelligence methods such as SHAP and Local Interpretable Model-agnostic Explanations have been applied to interpret fraud detection~\cite{thomas2026XAIDEFI}, anomaly detection~\cite{hasan2024detecting}, and price forecasting~\cite{indranil2023XAIDEFI}. In this work, we use model explanations to characterize the network structures associated with individual DeFi methods.

Our previous work~\cite{tovanich2025decoding} introduced ego-network motifs to classify DeFi transaction methods, while Wu et al.~\cite{wu2023mots} used motifs to represent transaction semantics.
We extend these approaches to complete transaction graphs to capture more complex DeFi methods, enrich graphlet features with account and token information, and assess their complementarity with event log count features. Differences in data and methods limit direct comparison with our previous work.

\section{Material and Methods}\label{sec:methods}

\subsection{Data}
We collect complete token transfer data for Ethereum transactions within blocks 20,000,000 to 20,019,999 (20,000 blocks) from the Alchemy Transfer API. 
For each transaction, we retrieve the transaction metadata and execution status from our own Ethereum node. We split the first 10,000 blocks into the training set and the remaining 10,000 into the test set.

Before constructing the transaction graphs, we apply the following preprocessing steps: (1) remove simple ETH transfers (i.e., \texttt{input = 0x}); (2) remove failed transactions during execution; (3) remove contract creation transactions; and (4) remove transfers with missing sender or recipient addresses. The resulting training and test datasets contain 590,589 and 666,342 transactions, respectively, totaling 1,932,808 and 1,909,794 token transfers.

The prediction target is the transaction method invoked by the transaction. We obtain decoded method labels from Etherscan as the ground truth. For transactions without a human-readable label, i.e., when the application binary interface is unavailable, we retain the four-byte function selector, which is represented by the first 10 characters of the input field, e.g., \texttt{0x0162e2d0}.
From the training set, we select the 30 most frequent methods: 24 have human-readable labels, while 6 are represented by function selectors. Together, these classes cover 73.78\% of all transactions, with class sizes ranging from 2,825 to 240,552. The full list of the 30 methods is given in Table~\ref{tab:app_method-performance} (Appendix~\ref{app:method-performance}).

\subsection{Token Transfer Network}

We represent each transaction as a directed, attributed multigraph: $G_{tx} = (V_{tx}, E_{tx})$, where $V_{tx}$ is the set of accounts involved in the transaction $tx$ and $E_{tx}$ is the set of token transfers. Multiple transfers between the same pair of accounts are retained when they involve different tokens.

\vspace{0.5em}

\noindent\textbf{Nodes.} Each node represents an Ethereum address assigned one of the following account types, which can be inferred directly from blockchain data:
\begin{itemize}
    \item \textit{Externally owned account}: an address controlled by a private key, primarily a user account (referred to as \emph{Account});
    \item \textit{Smart contract account}: an address containing deployed bytecode, possibly the protocol contract (referred to as \emph{Contract});
    \item \textit{Null address}: the zero address (\texttt{0x00000000...}) and a few other designated addresses used for token minting, burning, or system operations; and
    \item \textit{Token contract}: a smart contract address corresponding to a token's contract.
\end{itemize}

\vspace{0.5em}

\noindent\textbf{Edges.} Each edge is assigned a token type based on its transfer standard: (i) ETH, (ii) ERC-20 (fungible token), (iii) ERC-721 and (iv) ERC-1155, both used for non-fungible tokens (NFTs), or (v) non-standard NFTs (e.g., CryptoPunks and CryptoKitties).

In addition, we assign the 1,000 most frequently transferred tokens to a broader semantic category based on their economic function.
\begin{itemize}
    \item \textit{ETH/WETH}: native Ether and Wrapped Ether, grouped together since WETH is the ERC-20 representation of ETH;
    \item \textit{Cryptocurrency}: assets primarily used as transferable currencies or stores of value (stablecoins, fiat- or commodity-backed tokens, and meme coins), e.g., USDT, USDC, DAI, PAXG, SHIB, and PEPE;
    \item \textit{Protocol token}: native, governance, or utility tokens associated with DeFi protocols, e.g., BNB, LINK, AAVE, UNI, COMP, and ARB; and
    \item \textit{Synthetic asset}: tokens whose value derives from another asset (wrapped cross-chain assets, liquid-staking tokens, interest-bearing tokens), e.g., WBTC, stETH, wstETH, cbBTC, cETH, and cUSDC.
\end{itemize}

We deliberately avoid categorizing tokens by associated DeFi protocol type (e.g., exchange, lending, aggregator), since such labels could cause the classifier to learn application identities rather than generalizable transaction structures.

\subsection{Graphlet Pattern Matching}
To characterize the local topology of each transaction network, we apply a directed graphlet decomposition based on the directed graphlet definitions introduced in Sarajlić et al.~\cite{sarajlic2016}. A directed graphlet is a small, connected directed graph that represents a recurring local interaction pattern. Each graphlet contains a set of node positions (orbits) that distinguish the structural roles.

For each transaction graph, we identify subgraphs isomorphic to the predefined directed graphlet patterns ($G0$–$G39$), preserving edge direction and assigning each account to its graphlet position ($0$–$128$). 
We follow the graphlet and node position numbering of~\cite{sarajlic2016}.
All graphlets considered in this study are shown in Fig.~\ref{fig:app_directed_graphlets} in the Appendix, adapted from Fig.~1 of \cite{sarajlic2016}.
Each matched transfer is assigned to an edge position defined by the source and destination positions in the graphlet.
%\textcolor{red}{All considered graphlets are displayed in Appendix Figure~\ref{fig:app_directed_graphlets} with illustration following Fig.~1 from \cite{sarajlic2016}.}

Large star-shaped graphs can produce a combinatorial number of isomorphic subgraphs. To avoid excessive computation, for transactions that match the elementary star-like graphlet, we count $G0$ directly rather than enumerating all isomorphic combinations, thereby preserving transaction-size information while preventing combinatorial explosion.

\subsection{Feature Extraction}
\label{sec:featureextraction}

Since a single account or transfer can participate in multiple graphlet matches, counting every isomorphic match can lead to substantial repeated counting. For graphlet count features, we count matches per pattern. For node- and edge-position features, we instead count distinct entities rather than the total number of isomorphic matches. The graphlet-based features contain four feature groups.

\begin{itemize}
    \item \textit{Graphlet pattern counts}: the number of matches per directed graphlet pattern, e.g., $G0$, $G1$, $G2$;
    \item \textit{Node position counts}: the number of distinct accounts in each graphlet node position, e.g., $0$, $1$, $2$;
    \item \textit{Node position with account type counts}: the number of distinct accounts per node position and account type, e.g., \emph{0\_Contract}, \emph{1\_Token}, \emph{26\_Address}; and
    \item \textit{Edge position with token type/category counts}: the number of distinct (\emph{sender position}, \emph{recipient position}, \emph{token type/category}) tuples per edge position, e.g., \emph{0\_1\_ETH/WETH}.
\end{itemize}

We do not construct joint features that combine account type, token type, and category to avoid an overly large and sparse feature space. We retain raw counts without normalization because the number of repeated structural patterns, participating accounts, and transfer volumes conveys information about a transaction's scale and complexity.

\subsection{Transaction Method Classification}

To determine whether transaction methods can be inferred from the topology of token transfer graphs and event log counts, we evaluate transaction method classification performance in three experiments: 
(1) the complete graphlet-based representation with an ablation study removing each of the type-based feature groups; (2) comparison with event log count features; and (3) combination of both feature types to examine whether they provide complementary information for transaction-method classification.

For each configuration, we use AutoGluon Tabular version 1.5.0, an automated machine learning library that trains multiple classification models and combines them via bagging, stacking, and weighted ensembling~\cite{erickson2020autogluon}. AutoGluon optimizes the macro-averaged F1-score on the validation set and evaluates on a held-out test set. Given that transaction method classes are highly imbalanced, we report macro-averaged F1, precision, recall, and the multiclass one-vs.-one area under the receiver operating characteristic curve (ROC-AUC).

\subsection{SHAP Analysis and Graphlet Importance}
\label{subsec:graphlet-shap}

We use SHAP to interpret the classification model by quantifying the contributions of individual features to its predictions~\cite{lundberg2017}.
Shapley values provide an additive decomposition of a model's output into a reference value (baseline) and individual feature contributions.
Formally, given a transaction $x$ and the model prediction probability $p_m(x)$ for method $m$, $p_m(x) = \phi_{0,m} + \sum_{j=1}^{k} \phi_{j,m}(x)$, where $k$ is the number of features, $\phi_{0,m}$ is the average prediction of the model with respect to a background dataset, and $\phi_{j,m}(x)$ denotes the SHAP value associated with feature $j$ for transaction $x$.

For a baseline dataset $\mathcal{B}$, we have $\phi_{0,m} = \mathbb{E}_{x\in \mathcal{B}}[p_m(x)]$. A positive SHAP value indicates that the corresponding feature increases the model output for method $m$ relative to the reference value, whereas a negative value decreases it.

\vspace{0.5em}

\noindent\textbf{Method-specific baseline.}
We analyze topological features that distinguish transactions processed by method $m$ from those processed by other methods.
Since the background dataset directly influences the interpretation of SHAP values~\cite{chen2022}, we propose a method-specific baseline composed of transactions that are unlikely to exhibit method $m$: $\mathcal{B}_m = \{ x : p_m(x) \leq \kappa_u\}$, where $\kappa_u$ is an upper probability bound.

Under this formulation, SHAP contributions describe how features shift the model output for $m$ relative to transactions with low predicted association with $m$. Because $\phi_{0,m}$ differs across methods, graphlet-signature magnitudes should be interpreted \emph{within}, rather than \emph{across}, methods.

\vspace{0.5em}

\noindent\textbf{Feature packing and explainability set.}
For the sake of parsimony and explainability, we aggregate SHAP contributions at the graphlet level rather than at the level of individual features. We use a feature packing method based on the SHAP additivity property~\cite{noharaetal2021}. Let $S_G$ be the set of features associated with graphlet $G$. The \emph{aggregated graphlet contribution} for transaction $x$ is $\Phi_{G,m}(x) = \sum_{j \in S_G} \phi_{j,m}(x)$. Since this aggregation preserves the additive property of SHAP, the sum of these contributions plus the baseline recovers the model's prediction. 

We apply the SHAP analysis to a method-specific explainability set $\mathcal{T}_m$ drawn from the test set: $\mathcal{T}_m = \{ x \in \mathcal{X}_\mathrm{test} : p_m(x) \geq \kappa_l\}$,
where $\kappa_l$ is a lower probability bound, with $|\mathcal{T}_m| = 1{,}000$ (or fewer if support is limited).

\vspace{0.5em}

\noindent\textbf{Graphlet importance score.}
For a given method $m$ and its respective explainability set $\mathcal{T}_m$, we define the \emph{directional} and \emph{strength} graphlet importance scores:
\begin{equation*}
    \Pi_{G,m}^{\mathrm{D}} = \frac{\sum_{x \in \mathcal{T}_m} \Phi_{G,m}(x)}{|\mathcal{T}_m|}, \qquad
    \Pi^{\mathrm{S}}_{G,m} = \frac{\sum_{x \in \mathcal{T}_m} |\Phi_{G,m}(x)|}{|\mathcal{T}_m|}.
\end{equation*}
A positive (negative) sign of $\Pi_{G,m}^{\mathrm{D}}$ indicates increased (decreased) predicted probability of $m$, whereas $\Pi_{G,m}^{\mathrm{S}}$ measures the absolute impact of graphlet $G$ on the prediction.

Both scores are independently normalized per method: $\tilde{\Pi}_{G,m}^{\mathrm{D}}$ is obtained by dividing $\Pi_{G,m}^{\mathrm{D}}$ by $\max_{G'}\Pi_{G',m}^{\mathrm{D}}$ for positive contributions and by $-\min_{G'}\Pi_{G',m}^{\mathrm{D}}$ for negative contributions, and $\tilde{\Pi}_{G,m}^{\mathrm{S}} = \Pi_{G,m}^{\mathrm{S}} / \max_{G'}\Pi_{G',m}^{\mathrm{S}}$. Therefore, $\tilde{\Pi}_{G,m}^{\mathrm{D}}$ and $\tilde{\Pi}_{G,m}^{\mathrm{S}}$ are bounded in $[-1,1]$ and $[0,1]$, respectively.

\section{Results}\label{sec:results}

\subsection{Transaction Method Classification}
    
We evaluate whether transaction methods can be inferred from the topology of the token transfer network. First, we conduct an ablation study to examine the contribution of the different graphlet-based feature groups.
Table~\ref{tab:classification} shows that \textit{all graphlets} achieve an F1-score of 0.693. Removing \textit{graphlet count} features results in only a marginal decrease (0.691), suggesting their structural information is largely captured by typed positional features.
In contrast, removing \textit{account type} features causes the largest drop (0.592) while removing \textit{token type} and \textit{token category} features results in smaller decreases (0.669 and 0.681). These results indicate that the types of entities in graphlet positions are important for distinguishing transaction methods.

\begin{table}[t]
\centering
\caption{Classification performance of graphlet-based, event log, and combined feature representations. The \# column reports the number of features or, for ablations, the number removed from the complete graphlet-based representation. Best value per metric within each block in \textbf{bold}. The node position count group (129 features) is retained throughout.}
\label{tab:classification}
\resizebox{\linewidth}{!}{%
\begin{tabular}{p{4cm}rccccp{3.5cm}}
\toprule
\textbf{Features} & \textbf{\#} & \textbf{F1-score} & \textbf{Precision} & \textbf{Recall} & \textbf{ROC-AUC} & \textbf{Best model} \\
\midrule
All graphlets & 1{,}680 & \textbf{0.693} & 0.790 & \textbf{0.699} & \textbf{0.984} & WeightedEnsemble\_L2 \\
No graphlet counts & $-$40 & 0.691 & 0.787 & 0.697 & \textbf{0.984} & WeightedEnsemble\_L2 \\
No account types & $-$505 & 0.592 & 0.727 & 0.599 & 0.969 & WeightedEnsemble\_L2 \\
No token types & $-$504 & 0.669 & \textbf{0.793} & 0.667 & 0.981 & WeightedEnsemble\_L2 \\
No token categories & $-$502 & 0.681 & 0.748 & 0.691 & 0.983 & WeightedEnsemble\_L2 \\
No token types \& categories & $-$1{,}006 & 0.638 & 0.746 & 0.632 & 0.977 & NeuralNetFastAI \\
\midrule
Top 100 events & 100 & 0.750 & 0.835 & 0.742 & 0.984 & WeightedEnsemble\_L2 \\
Top 1{,}000 events & 1{,}000 & 0.791 & 0.853 & 0.784 & 0.988 & NeuralNetTorch \\
All events & 4{,}693 & \textbf{0.793} & \textbf{0.862} & \textbf{0.787} & \textbf{0.989} & WeightedEnsemble\_L2 \\
\midrule
Combined graphlet features and top 1{,}000 events & 2{,}680 & \textbf{0.918} & \textbf{0.953} & \textbf{0.918} & \textbf{0.998} & WeightedEnsemble\_L2 \\
\bottomrule
\end{tabular}%
}
\end{table}

Second, we compare the graphlet representation with event log count features. Table~\ref{tab:classification} shows that event log features outperform graphlet features, providing a more direct indication of the invoked method. The \textit{top 100 events} achieve an F1-score of 0.750, while the \textit{top 1,000 events} reach 0.791. Using all 4,693 events provides only a marginal improvement (0.793), indicating that the top 1,000 events capture most of the predictive information. 
Combining \textit{all graphlets} with the \textit{top 1,000 event logs} achieves the strongest overall performance (F1 = 0.918, precision = 0.953, recall = 0.918), showing the two representations provide complementary information. Event logs capture emitted operations during execution, while graphlet features encode the structure of token transfers and interactions among accounts and tokens.

%\textcolor{red}{We present classification score contribution at the method level in case of the top 30 DeFi methods in Appendix Table~\ref{tab:app_method-performance}.}

\begin{figure}[t]
    \centering
    \includegraphics[width=\linewidth]{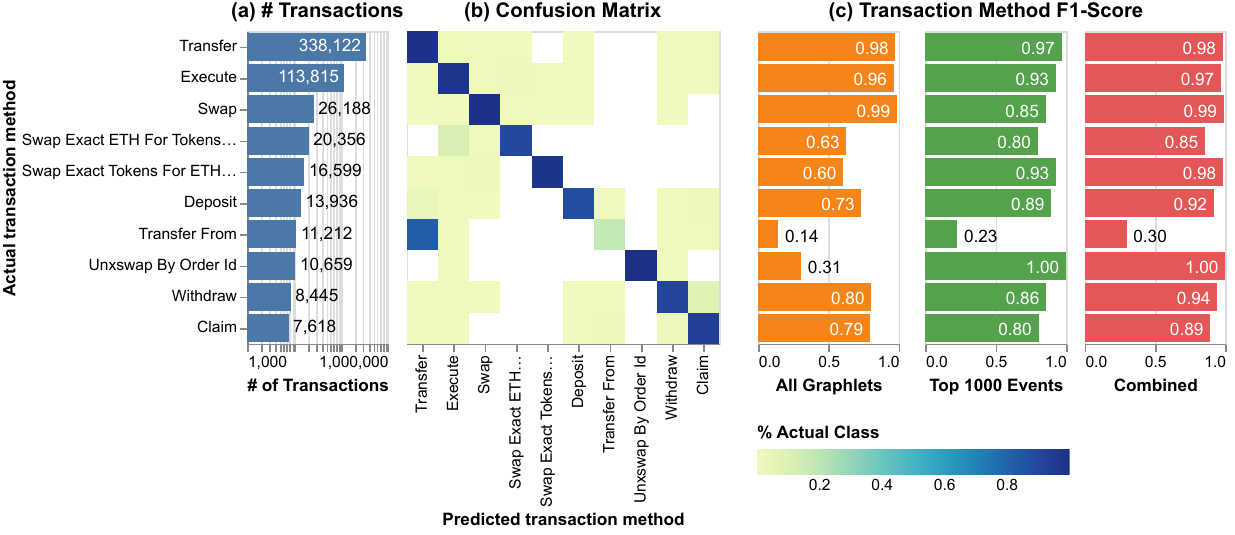}
    \caption{Classification results for the top-10 transaction methods: (a) number of transactions per method on the test set, (b) recall-normalized confusion matrix of the combined model, and (c) per-method F1-scores for graphlet-based, event log count, and combined feature models. Panel (a) uses a logarithmic axis. Rows of panel (b) sum to one, so diagonal entries give recall.}
    \label{fig:confusion}
\end{figure}

We further analyze the recall-normalized confusion matrix on the top 10 methods for the combined model in Fig.~\ref{fig:confusion}(b). 
Each row corresponds to an actual transaction method and sums to 1, while the columns represent predicted methods. 
Almost all methods are correctly classified, as indicated by the strong diagonal concentration, while errors mainly occur between methods with closely related functionality or overlapping execution behavior.

The two outstanding confusions are 82\% of \texttt{Transfer From} misclassified as \texttt{Transfer}, followed by 74\% of \texttt{Swap Exact ETH For Tokens} (which is outside the top-10 list) misclassified as \texttt{Swap Exact ETH For Tokens Supporting Fee On Transfer Tokens}. The complete confusion matrix over all 30 methods, together with further confusion patterns, is provided in Fig.~\ref{fig:app_confusion} (Appendix~\ref{app:confusion}).
%\textcolor{red}{ We present the confusion matrix in case of the top 30 DeFi methods in Appendix Figure~\ref{fig:app_confusion}.}

Finally, we examine method-level performance to study how the two feature representations contribute to classification performance. Fig.~\ref{fig:confusion}(c) compares the F1-scores across the top 10 methods, showing that individual methods rely on different signals.
Graphlet features outperform event logs for \texttt{Swap} (0.986 vs.\ 0.855) and \texttt{Execute} (0.964 vs.\ 0.927), suggesting that these methods exhibit distinctive transfer structures. 
Conversely, event logs perform better for the majority of methods, such as \texttt{Unxswap By Order Id} (1.000 vs.\ 0.307), \texttt{Swap Exact Tokens For ETH Supporting Fee On Transfer Tokens} (0.929 vs.\break 0.604), \texttt{Deposit} (0.895 vs.\ 0.735), and \texttt{Withdraw} (0.858 vs.\ 0.805). Precision, recall, and F1-score for all 30 methods are reported in Table~\ref{tab:app_method-performance} (Appendix~\ref{app:method-performance}).
%Combining both representations yields the best performance, either maintaining the performance of the stronger feature group or further improving classification for many methods.
Combining both representations yields the best performance, either maintaining the performance of the stronger feature group or further improving classification for 29 of the 30 methods. The only exception is \texttt{Deposit ETH}, which falls outside the top 10 methods, with an F1-score of 0.903 for the combined representation compared to 0.989 for event logs alone.

\subsection{SHAP Analysis and Graphlet Importance}
We analyze SHAP contributions at both the feature and graphlet levels, with heuristically chosen thresholds $\kappa_u=0.2$ and $\kappa_l=0.8$ for the method-specific reference backgrounds $\mathcal{B}_m$ and explainability sets $\mathcal{T}_m$.
We apply the permutation explainer to the second-best model, \emph{NeuralNetFastAI} (F1-score $=0.91769$), AutoGluon's implementation of fast.ai's tabular neural network~\cite{autogluon_fastai_nn}. Its performance is close to that of the best model, \emph{WeightedEnsemble\_L2}, and a single model is simpler to explain than an ensemble.

\begin{table}[t]
\centering
\caption{Top 10 labeled transaction methods with their GPR and 10 most important features by mean absolute SHAP value. Graphlet-based features are shown in bold, with importance values in parentheses. See Section~\ref{sec:featureextraction} for feature naming conventions. Results for all 30 methods are given in Tables~\ref{tab:app_top_features_by_method} and~\ref{tab:app_top_features_by_method_ext} (Appendix~\ref{app:shap-features}).
}
\label{tab:top_features_by_method}
\scriptsize
\begin{tabularx}{\textwidth}{>{\raggedright\arraybackslash}p{2.5cm} p{1.1cm} X}
\toprule
\textbf{Method} & \textbf{GPR (\%)} & \textbf{Top 10 most important features} \\
\midrule
Transfer & 67.20 & \textbf{0\_Contract (0.25)}, Approval (0.09), Swap\_4 (0.08), Sync (0.08), Swap\_5 (0.06), \textbf{0\_1\_ETH/WETH (0.03)}, \textbf{1\_Contract (0.03)}, Deposit\_7 (0.03), Permit (0.02), \textbf{14\_Address (0.02)} \\
Execute & 62.28 & Permit (0.11), Swap\_3 (0.08), \textbf{26\_Address (0.07)}, \textbf{1\_Contract (0.05)}, TransformedERC20 (0.05), OrderRecord (0.04), BridgeFill (0.03), \textbf{0\_Contract (0.02)}, \textbf{13\_Address (0.02)}, \textbf{16\_Address (0.02)} \\
Swap & 58.65 & Swap\_3 (0.35), \textbf{26\_Address (0.07)}, Permit (0.05), \textbf{0\_Contract (0.05)}, TransformedERC20 (0.03), \textbf{1\_Contract (0.03)}, \textbf{16\_Address (0.03)}, \textbf{13\_Address (0.02)}, ClientData (0.02), 0 (0.02) \\
Swap Exact ETH For Tokens ... & 74.99 & \textbf{1\_Contract (0.09)}, \textbf{0\_Contract (0.07)}, \textbf{0\_1\_Cryptocurrency (0.06)}, Deposit\_7 (0.06), \textbf{0\_1\_ETH (0.05)}, Permit (0.03), \textbf{28\_Address (0.02)}, Swap\_4 (0.02), \textbf{51\_52\_erc20 (0.02)}, \textbf{2\_Address (0.02)} \\
Swap Exact Tokens For ETH ... & 68.61 & Swap\_4 (0.06), Permit (0.05), Swap\_5 (0.05), \textbf{0\_1\_ETH (0.05)}, \textbf{0\_Contract (0.05)}, \textbf{26\_Address (0.04)}, \textbf{1\_Contract (0.04)}, TransformedERC20 (0.04), \textbf{53\_Address (0.03)}, Withdrawal (0.03) \\
Deposit & 62.44 & \textbf{0\_Contract (0.11)}, Deposit\_7 (0.10), \textbf{1\_Contract (0.06)}, Swap\_5 (0.05), \textbf{4\_Address (0.05)}, Deposit\_5 (0.04), LogDeposit (0.04), \textbf{14\_Address (0.03)}, Swap\_4 (0.03), Permit (0.03) \\
Transfer From & 79.01 & \textbf{0\_1\_ETH/WETH (0.11)}, \textbf{0\_Contract (0.09)}, \textbf{0\_1\_erc721 (0.07)}, Swap\_5 (0.06), \textbf{5\_Address (0.06)}, Approval (0.04), \textbf{1\_Contract (0.04)}, \textbf{2\_Address (0.04)}, Swap\_4 (0.03), Transfer\_3 (0.03) \\
Unxswap By Order Id & 29.44 & OrderRecord (0.63), Swap\_4 (0.08), Sync (0.03), Permit (0.03), \textbf{0\_Contract (0.03)}, Swap\_5 (0.02), \textbf{0\_1\_Protocol (0.02)}, \textbf{1\_Contract (0.01)}, Swap\_3 (0.01), TransformedERC20 (0.01) \\
Withdraw & 61.22 & \textbf{1\_Contract (0.08)}, \textbf{4\_Address (0.07)}, \textbf{0\_1\_Cryptocurrency (0.06)}, \textbf{2\_Address (0.05)}, Withdrawn\_3 (0.04), Swap\_5 (0.03), Withdraw (0.03), WithdrawFromLockupStream (0.03), Withdrawal (0.03), Withdraw\_5 (0.03) \\
Claim & 53.83 & \textbf{1\_Contract (0.11)}, TokenSent (0.05), 0x9d...a305 (0.04), Claimed\_6 (0.04), 0x46...911f (0.03), Swap\_5 (0.03), 0xa3...7852 (0.03), \textbf{0\_1\_Cryptocurrency (0.03)}, Claimed\_9 (0.03), \textbf{4\_Address (0.03)} \\
\bottomrule
\end{tabularx}
\end{table}

\vspace{0.5em}

\noindent\textbf{Feature-level SHAP analysis.}
We examine the contribution of graphlet-based and event-log features to the classifier's predictions.
For each of the top 10 methods, Table~\ref{tab:top_features_by_method} reports the graphlet participation ratio (GPR), the ratio of cumulative absolute SHAP values of graphlet-based features to those of all features. Graphlet-based features dominate most methods, with GPR above 50\%. The highest ratio is for \texttt{Transfer From} (79.01\%).
The exception is \texttt{Unxswap By Order Id} (29.44\%), whose predictions rely primarily on event log features. The corresponding results for all 30 methods, including the unlabeled function selectors, are given in Tables~\ref{tab:app_top_features_by_method} and~\ref{tab:app_top_features_by_method_ext} (Appendix~\ref{app:shap-features}).

The top 10 features with the largest mean absolute SHAP contribution (Table~\ref{tab:top_features_by_method}) combine both graphlet-based and event-log features. Among graphlet-based features, the most important ones involve \emph{node position with account type} (e.g., \emph{0\_Contract}, \emph{1\_Contract}, \emph{26\_Address}). We also observe \emph{edge position with token type/category} (e.g., \emph{0\_1\_ETH}, \emph{0\_1\_Cryptocurrency}), especially for transfer-related and swap-related methods.

Specific graphlet features also inform how local topology can characterize DeFi methods.
For example, \texttt{Deposit} is associated with \emph{0\_Contract}, whereas \texttt{Withdraw} is conversely characterized by \emph{1\_Contract}, reflecting opposite asset-flow directions relative to the contract account.
This suggests that predictive power depends not only on graphlet topology, but also on the semantic types of accounts and tokens in specific structural positions.

%\textcolor{red}{Tables~\ref{tab:app_top_features_by_method} and \ref{tab:app_top_features_by_method_ext} from Appendix present the case of top 30 methods and provide measures of graphlet dominance (\% graphlet) defined as the proportion of transactions in $\mathcal{T}_m$ presenting higher cumulative absolute SHAP values for graphlet features than event-log features.}

\vspace{0.5em}

\begin{figure}[t]
    \centering
        \includegraphics[width=\linewidth]{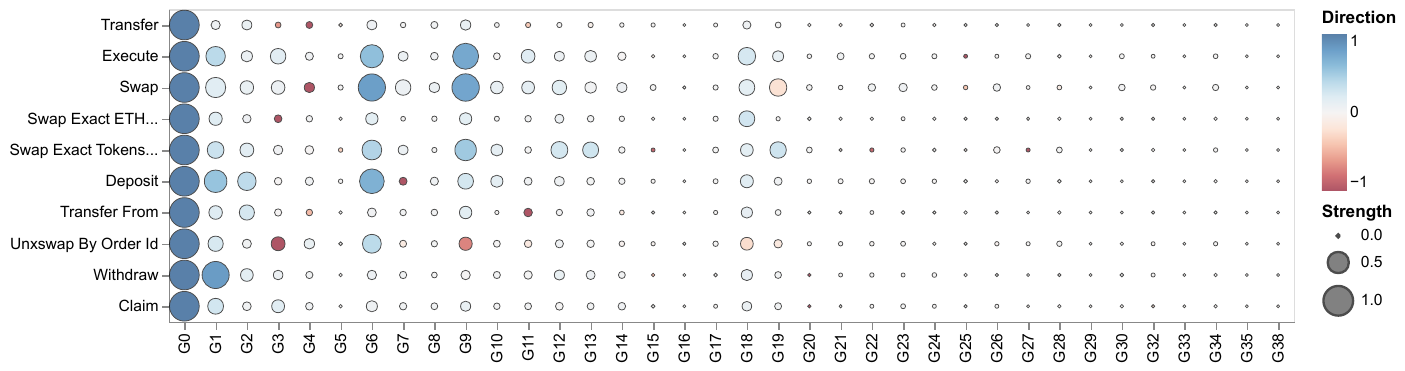}
    \caption{Method-specific graphlet signatures from aggregated SHAP contributions. 
    Rows are methods, columns are directed graphlets. Circle area gives the normalized strength $\tilde{\Pi}_{G,m}^{\mathrm{S}}$ and color the normalized direction $\tilde{\Pi}_{G,m}^{\mathrm{D}}$, blue positive and red negative relative to the baseline $\mathcal{B}_m$.
    $G31$, $G36$, $G37$, and $G39$ are absent from our data.}
    \label{fig:heatmap}
\end{figure}

\noindent\textbf{Graphlet signatures.}
We aggregate feature contributions to identify which graphlet characterizes each method independently of specific attribute combinations. For every graphlet $G$ and method $m$, we report the normalized directional ($\tilde{\Pi}_{G,m}^{\mathrm{D}}$) and strength ($\tilde{\Pi}_{G,m}^{\mathrm{S}}$) importance in Fig.~\ref{fig:heatmap}.

In terms of the graphlet strengths, the simplest graphlet, $G0$, contributes strongly to predictions across all top-10 methods and is dominant for \texttt{Transfer}, consistent with its predominantly dyadic token-transfer structure. 
The small graphlets $G1$--$G3$ also exhibit high strengths across many methods. This aligns with swap-related methods, which typically involve chains of token transfers and intermediate contracts. Similar patterns appear for \texttt{Deposit}, \texttt{Withdraw}, and \texttt{Execute}.

For several methods, the signatures extend beyond small graphlets. \texttt{Swap} and its variants, as well as \texttt{Execute}, show strong contributions from $G6$, $G9$, $G12$, $G13$, $G18$, and $G19$, suggesting that some DeFi operations involve multi-step interactions among multiple accounts and transfers beyond simple dyadic structures. The heatmap indicating the graphlet signatures for all 30 methods are shown in Fig.~\ref{fig:app_heatmap} (Appendix~\ref{app:signatures}).
%\textcolor{red}{We present in the Appendix Figure~\ref{fig:app_heatmap} the heatmap in case of the top 30 most important DeFi methods.}

\section{Conclusion}\label{sec4}

We investigated whether token-transfer topology can characterize DeFi transaction methods using directed graphlets enriched with account and token information. Graphlet features achieve a macro F1-score of 0.693, compared with 0.793 for event logs, while combining both representations substantially improves performance to 0.918. Method-level results further show that some operations are better characterized by transfer structures and others by emitted events.

SHAP analysis reveals that graphlet-based features contribute to predictions for most analyzed methods. We identify method-specific structural signatures, ranging from simple transfer patterns to more complex multi-account interactions. These results show that graphlets provide both predictive and interpretable representations of DeFi transaction behavior.

Our test window directly follows the training window, so generalization to distant periods, unseen contracts, and other protocols remains untested. Future work will evaluate larger datasets, benchmark graphlet features against baseline models such as standard network metrics, and compare SHAP values with tree-based explanations and non-heuristic thresholding strategies. As DeFi transactions increasingly combine complex activities such as swaps, lending, flash loans, and arbitrage, we aim to move beyond transaction-level classification to decompose token transfer networks into their underlying operations. We also wish to investigate whether structural positions can reveal the functional roles of accounts and smart contracts within DeFi protocols.

\section*{Acknowledgement}
We sincerely thank Thummanoon Kunanuntakij for his initial work, which helped shape the framing of this paper, and for his thoughtful discussions and feedback throughout the project.

%
% ---- Bibliography ----
%
\bibliographystyle{spmpsci}
\bibliography{main.bib}

@inproceedings{kosyfaki2019flow,
author = {Kosyfaki, Chrysanthi and Mamoulis, Nikos and Pitoura, Evaggelia and Tsaparas, Panayiotis},
title = {Flow Motifs in Interaction Networks},
booktitle = {Proceedings of the 22nd International Conference on Extending Database Technology (EDBT)},
year = {2019},
pages = {241--252}
}

@article{indranil2023XAIDEFI,
title = {Prediction and interpretation of daily {NFT} and {DeFi} prices dynamics: Inspection through ensemble machine learning \& {XAI}},
journal = {International Review of Financial Analysis},
volume = {87},
pages = {102558},
year = {2023},
issn = {1057-5219},
doi = {10.1016/j.irfa.2023.102558},
url = {https://www.sciencedirect.com/science/article/pii/S1057521923000741},
author = {Indranil Ghosh and Esteban Alfaro-Cortés and Matías Gámez and Noelia García-Rubio}
}

@inproceedings{thomas2026XAIDEFI,
  author={Thomas, Rino and K.K, Savitha},
  booktitle={2026 Sixth International Conference on Advances in Electrical, Computing, Communications and Sustainable Technologies (ICAECT)}, 
  title={Pre-Transaction Fraud Risk Prediction in {DeFi} Using Explainable {AI}}, 
  year={2026},
  volume={},
  number={},
  pages={1-6},
  doi={10.1109/ICAECT68478.2026.11426021}
}

@article{ma2025DefiFraud,
  author={Ma, Wei and Shi, Junjie and Qiu, Jiaxi and Wu, Cong and Chen, Jing and Jiang, Lingxiao and Liu, Shangqing and Liu, Yang and Xiang, Yang},
  journal={IEEE Transactions on Information Forensics and Security}, 
  title={Detecting {DeFi} Fraud With a Graph-Transformer Language Model}, 
  year={2025},
  volume={20},
  number={},
  pages={10051-10065},
  doi={10.1109/TIFS.2025.3612184}
}

@inproceedings{rudrawar2025DeFiFraud,
  author={Rudrawar, Harsh and Jakhete, Sumitra A. and Somani, Dhanashree},
  booktitle={2025 IEEE International Conference on Blockchain and Distributed Systems Security (ICBDS)}, 
  title={Fraud Detection in Decentralized Finance ({DeFi}) Transactions using {ML} and Smart Contracts}, 
  year={2025},
  volume={},
  number={},
  pages={1-6},
  doi={10.1109/ICBDS67396.2025.11377130}
}

@article{tovanich2023fingerprinting,
  author  = {Tovanich, Natkamon and Cazabet, R{\'e}my},
  title   = {Fingerprinting {Bitcoin} entities using money flow representation learning},
  journal = {Applied Network Science},
  volume  = {8},
  number  = {1},
  pages   = {63},
  year    = {2023},
  doi     = {10.1007/s41109-023-00591-2}
}

@article{wu2022detecting,
  author  = {Wu, Jiajing and Liu, Jieli and Chen, Weili and Huang, Huawei and Zheng, Zibin and Zhang, Yan},
  title   = {Detecting Mixing Services via Mining {Bitcoin} Transaction Network With Hybrid Motifs},
  journal = {IEEE Transactions on Systems, Man, and Cybernetics: Systems},
  volume  = {52},
  number  = {4},
  pages   = {2237--2249},
  year    = {2022},
  doi     = {10.1109/TSMC.2021.3049278}
}

@article{aoyama2022cryptoasset,
  author  = {Aoyama, Hideaki and Fujiwara, Yoshi and Hidaka, Yoshimasa and Ikeda, Yuichi},
  title   = {Cryptoasset networks: Flows and regular players in {Bitcoin} and {XRP}},
  journal = {PLOS ONE},
  volume  = {17},
  number  = {8},
  pages   = {e0273068},
  year    = {2022},
  doi     = {10.1371/journal.pone.0273068}
}

@incollection{tovanich2025decoding,
  author    = {Tovanich, Natkamon and Coquidé, Célestin and Cazabet, Rémy},
  title     = {Decoding Decentralized Finance Transactions Through Ego Network Motif Mining},
  booktitle = {Complex Networks \& Their Applications XIII},
  series    = {Studies in Computational Intelligence},
  volume    = {1188},
  pages     = {150--162},
  year      = {2025},
  publisher = {Springer},
  address   = {Cham},
  doi       = {10.1007/978-3-031-82431-9_13}
}

@article{arnold2024insights,
  author  = {Arnold, Naomi A. and Zhong, Peijie and Ba, Cheick Tidiane and Steer, Ben and Mondragon, Raul and Cuadrado, Felix and Lambiotte, Renaud and Clegg, Richard G.},
  title   = {Insights and caveats from mining local and global temporal motifs in cryptocurrency transaction networks},
  journal = {Scientific Reports},
  volume  = {14},
  pages   = {26569},
  year    = {2024},
  doi     = {10.1038/s41598-024-75348-7}
}

@article{chen2022,
  author  = {Chen, Hugh and Lundberg, Scott M. and Lee, Su-In},
  title   = {Explaining a series of models by propagating {Shapley} values},
  journal = {Nature Communications},
  volume  = {13},
  pages   = {4512},
  year    = {2022},
  doi     = {10.1038/s41467-022-31384-3}
}

@inproceedings{lundberg2017,
  author    = {Lundberg, Scott M. and Lee, Su-In},
  title     = {A Unified Approach to Interpreting Model Predictions},
  booktitle = {Advances in Neural Information Processing Systems},
  volume    = {30},
  pages     = {4768--4777},
  year      = {2017},
  publisher = {Curran Associates, Inc.},
  url       = {https://dl.acm.org/doi/10.5555/3295222.3295230}
}

@article{noharaetal2021,
  author  = {Nohara, Yasunobu and Matsumoto, Koutarou and Soejima, Hidehisa and Nakashima, Naoki},
  title   = {Explanation of machine learning models using {Shapley} additive explanation and application for real data in hospital},
  journal = {Computer Methods and Programs in Biomedicine},
  volume  = {214},
  pages   = {106584},
  year    = {2022},
  doi     = {10.1016/j.cmpb.2021.106584}
}

@article{milo2002,
author = {R. Milo  and S. Shen-Orr  and S. Itzkovitz  and N. Kashtan  and D. Chklovskii  and U. Alon },
title = {Network Motifs: Simple Building Blocks of Complex Networks},
journal = {Science},
volume = {298},
number = {5594},
pages = {824-827},
year = {2002},
doi = {10.1126/science.298.5594.824},
URL = {https://www.science.org/doi/abs/10.1126/science.298.5594.824},
eprint = {https://www.science.org/doi/pdf/10.1126/science.298.5594.824}}

@article{sarajlic2016,
  author  = {Sarajli{\'c}, Anida and Malod-Dognin, Noël and Yavero{\u{g}}lu, {\"O}mer Nebil and Pržulj, Nataša},
  title   = {Graphlet-based Characterization of Directed Networks},
  journal = {Scientific Reports},
  volume  = {6},
  pages   = {35098},
  year    = {2016},
  doi     = {10.1038/srep35098},
  url     = {https://doi.org/10.1038/srep35098},
  issn    = {2045-2322},
  publisher = {Nature Publishing Group}
  }

@misc{erickson2020autogluon,
  author       = {Erickson, Nick and Mueller, Jonas and Shirkov, Alexander and Zhang, Hang and Larroy, Pedro and Li, Mu and Smola, Alexander},
  title        = {{AutoGluon-Tabular}: Robust and Accurate {AutoML} for Structured Data},
  year         = {2020},
  howpublished = {arXiv preprint arXiv:2003.06505},
  doi          = {10.48550/arXiv.2003.06505},
  url          = {https://arxiv.org/abs/2003.06505}
}

@article{przulj2007,
    author = {Pržulj, Nataša},
    title = {Biological network comparison using graphlet degree distribution},
    journal = {Bioinformatics},
    volume = {23},
    number = {2},
    pages = {e177-e183},
    year = {2007},
    issn = {1367-4803},
    doi = {10.1093/bioinformatics/btl301},
    url = {https://doi.org/10.1093/bioinformatics/btl301},
    eprint = {https://academic.oup.com/bioinformatics/article-pdf/23/2/e177/49820586/bioinformatics_23_2_e177.pdf},
}

@article{kitzler2023,
  author    = {Kitzler, Stefan and Victor, Friedhelm and Saggese, Pietro and Haslhofer, Bernhard},
  title     = {Disentangling Decentralized Finance ({DeFi}) Compositions},
  journal   = {ACM Transactions on the Web},
  volume    = {17},
  number    = {2},
  pages     = {10:1--10:26},
  year      = {2023},
  publisher = {Association for Computing Machinery},
  issn      = {1559-1131},
  doi       = {10.1145/3532857},
  url       = {https://doi.org/10.1145/3532857}
}

@article{azad2025machine,
  title={Machine learning for blockchain data analysis: Progress and opportunities},
  author={Azad, Poupak and Akcora, Cuneyt Gurcan and Khan, Arijit},
  journal={Distributed Ledger Technologies: Research and Practice},
  volume={5},
  number={1},
  pages={1--27},
  year={2025},
  doi={10.1145/3728474}
}

@article{hasan2024detecting,
  title={Detecting anomalies in blockchain transactions using machine learning classifiers and explainability analysis},
  author={Hasan, Mohammad and Rahman, Mohammad Shahriar and Janicke, Helge and Sarker, Iqbal H},
  journal={Blockchain: Research and Applications},
  volume={5},
  number={3},
  pages={100207},
  year={2024},
  doi={10.1016/j.bcra.2024.100207}
}

@misc{autogluon_fastai_nn,
  author       = {{AutoGluon Contributors}},
  title        = {{AutoGluon Tabular Neural Network with FastAI Implementation}},
  year         = {2025},
  howpublished = {\url{https://github.com/autogluon/autogluon/blob/v1.5.0/tabular/src/autogluon/tabular/models/fastainn/tabular_nn_fastai.py}},
  note         = {AutoGluon 1.5.0 Documentation, Accessed: 2026-10-06}
}

@inproceedings{wang2023phishing,
  author    = {Yunfei Wang and Hao Wang and Xiaozhen Lu
               and Lu Zhou and Liang Liu},
  title     = {Detecting {Ethereum} Phishing Scams with Temporal Motif Features of Subgraph},
  booktitle = {2023 IEEE Symposium on Computers and Communications (ISCC)},
  pages     = {631--636},
  publisher = {IEEE},
  year      = {2023},
  doi       = {10.1109/ISCC58397.2023.10218023}
}

@inproceedings{wu2023mots,
  author    = {Zhiying Wu and Jieli Liu and Jiajing Wu and Zibin Zheng and Xiapu Luo and Ting Chen},
  title     = {Know Your Transactions: Real-time and Generic Transaction Semantic Representation on Blockchain \& {Web3} Ecosystem},
  booktitle = {Proceedings of the ACM Web Conference 2023},
  pages     = {1918--1927},
  year      = {2023},
  publisher = {ACM},
  doi       = {10.1145/3543507.3583537}
}

@inproceedings{akcora2020bitcoinheist,
  author    = {Cuneyt Gurcan Akcora and Yitao Li
               and Yulia R. Gel and Murat Kantarcioglu},
  title     = {{BitcoinHeist}: Topological Data Analysis for Ransomware Prediction on the {Bitcoin} Blockchain},
  booktitle = {Proceedings of the Twenty-Ninth International Joint Conference on Artificial Intelligence (IJCAI)},
  pages     = {4439--4445},
  year      = {2020},
  doi       = {10.24963/ijcai.2020/612}
}

@article{liu2024fishing,
  author  = {Jieli Liu and Jinze Chen and Jiajing Wu and Zhiying Wu and Junyuan Fang and Zibin Zheng},
  title   = {Fishing for Fraudsters: Uncovering {Ethereum} Phishing Gangs With Blockchain Data},
  journal = {IEEE Transactions on Information Forensics and Security},
  volume  = {19},
  pages   = {3038--3050},
  year    = {2024},
  doi     = {10.1109/TIFS.2024.3359000}
}
\clearpage
\section*{Appendix}
%Renumbering the sections, tables and figures reference indiex.
\setcounter{figure}{0}
\renewcommand{\thefigure}{A\arabic{figure}}

% Appendix table numbering: Table A1, Table A2, ...
\setcounter{table}{0}
\renewcommand{\thetable}{A\arabic{table}}
%Add a paragraph that describes a figure illustrating the considered graphlet patterns inspired by the Sarajlic et al. 2016
%\section{Results}\label{sec:app_results}

\appendix

This appendix provides the supplementary results referred to in the main text: the directed graphlet patterns (Appendix~\ref{app:graphlets}), the full confusion matrix and the per-method scores for all 30 transaction methods (Appendix~\ref{app:classification}), and the complete SHAP-based feature and graphlet analyses (Appendix~\ref{app:shap}).

\section{Directed Graphlets}
\label{app:graphlets} Fig.~\ref{fig:app_directed_graphlets} shows the 40 directed graphlets $G0$--$G39$ used to decompose the transaction graphs, adapted from Fig.~1 of~\cite{sarajlic2016}, whose graphlet and node position numbering we follow. These identifiers are the ones used for the graphlet pattern counts and node positions (Section~\ref{sec:featureextraction}) and for the graphlet signatures in Fig.~\ref{fig:app_heatmap}. Graphlets $G31$, $G36$, $G37$, and $G39$ do not occur in our data.
\input{graphlets_tikz}

\section{Transaction Method Classification}
\label{app:classification}
\subsection{Confusion Matrix of the Combined Model}
\label{app:confusion}

\begin{figure}
    \centering
    \includegraphics[width=\linewidth]{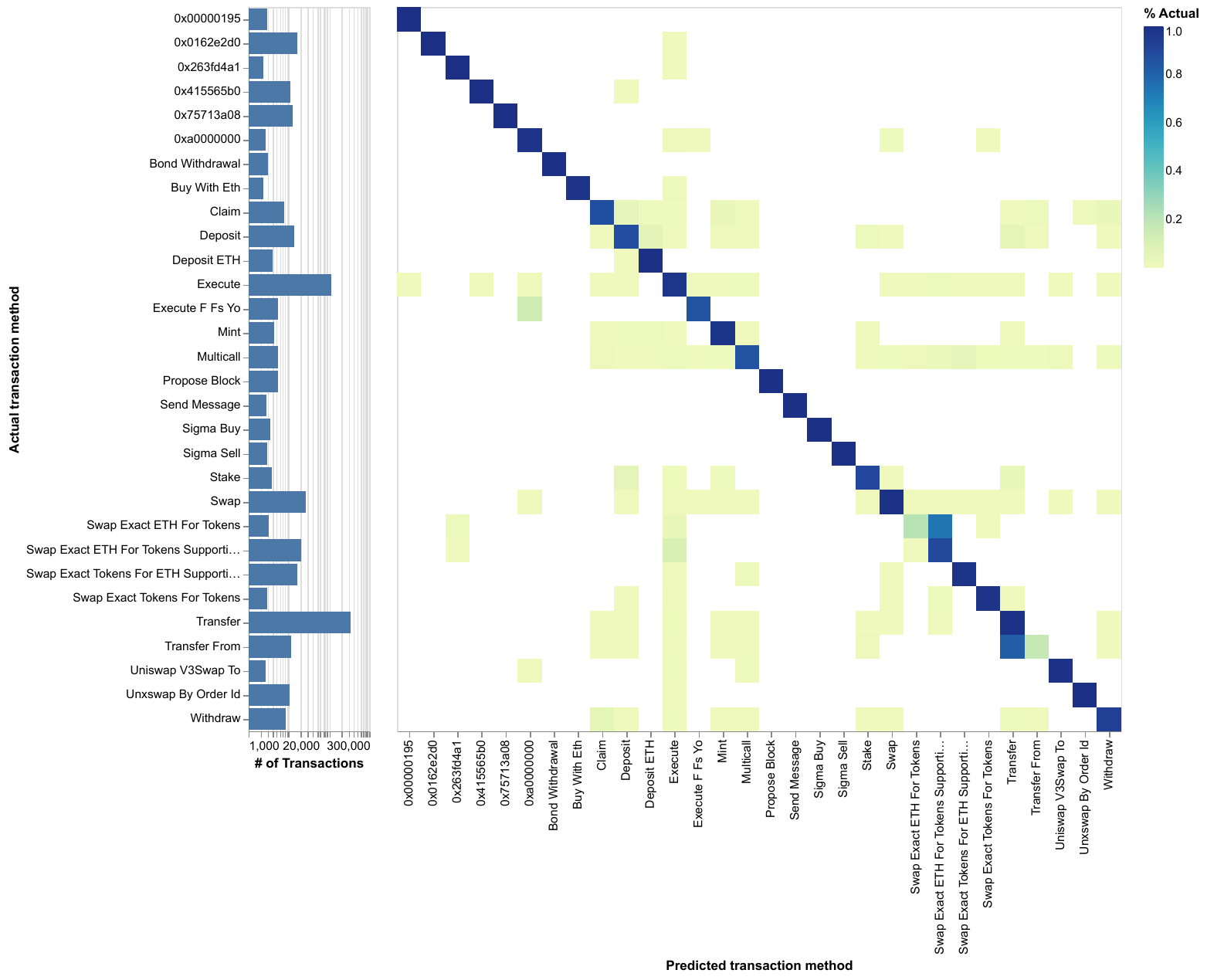}
    \caption{Recall-normalized confusion matrix for the combined graphlet-based and event-log features. Rows represent actual methods and columns represent predicted methods.}
    \label{fig:app_confusion}
\end{figure}

Figure~\ref{fig:app_confusion} presents the recall-normalized confusion matrix for the combined graphlet and event-log representation. Each row represents an actual transaction method, and the rows sum to 1. Consequently, diagonal entries correspond to method-level recall, while off-diagonal entries indicate the methods to which incorrectly classified transactions are assigned.

%\subsubsection{Confusion Matrix} The confusion matrix, as depicted in Figure~\ref{fig:confusion}, shows the recall-normalized confusion matrix for the combined representation. Most methods are classified correctly, as indicated by the strong diagonal. The largest confusion occurs between \texttt{Transfer From} and \texttt{Transfer}, with 81.8\% of \texttt{Transfer From} transactions assigned to \texttt{Transfer}. Similarly, 74.0\% of \texttt{Swap Exact ETH For Tokens} transactions are classified as \texttt{Swap Exact ETH For Tokens Supporting Fee On Transfer Tokens}. These errors occur between methods with closely related functionality, suggesting that they produce similar transfer structures and event profiles.

Most methods are classified correctly, as indicated by the strong diagonal concentration. Nevertheless, several confusions remain. The largest occurs between \texttt{Transfer From} and \texttt{Transfer}, with 81.8\% of \texttt{Transfer From} transactions classified as \texttt{Transfer}. Both methods represent token transfer operations and can produce closely related transfer structures and event profiles.
%Their distinction may depend primarily on the initiating account and allowance mechanism rather than on the resulting token flow. 

A second major confusion occurs between \texttt{Swap Exact ETH For Tokens} and \texttt{Swap Exact ETH For Tokens Supporting Fee On Transfer Tokens}. In total, 74.0\% of transactions associated with the former are classified as the supporting-fee-on-transfer variant. These methods perform closely related swap operations and differ mainly in how they handle tokens that charge fees during transfer.
%This implementation-level distinction may not always be observable from the extracted graphlet and event-count features. 

Other top misclassifications, although much less frequent, include classifying \texttt{Execute F Fs Yo} as \texttt{0xa0000000} (14.3\%) and \texttt{Swap Exact ETH For Tokens Supporting Fee On Transfer Tokens} as \texttt{Execute} (10.5\%).
Smaller confusion patterns include \texttt{Deposit} as \texttt{Deposit ETH} (5.6\%) or \texttt{Transfer} (5.1\%), \texttt{Withdraw} as \texttt{Claim} (5.3\%), and \texttt{Stake} as \texttt{Deposit} (4.9\%). 
%These errors generally occur between methods with related execution behavior or overlapping token transfer patterns.

These errors provide evidence that the remaining misclassifications are not arbitrary. Instead, they frequently occur between methods with closely related functionality or overlapping execution behavior.

\begin{table}[t]
    \centering
    \caption{Per-method classification performance for the graphlet-based representation (\textit{Graph.}), the top 1,000 event log count features (\textit{Events}), and their combined representation (\textit{Comb.}). The best result for each method and metric is shown in \textbf{bold}.} 
    \label{tab:app_method-performance} 
    \resizebox{\linewidth}{!}{% 
    \begin{tabular}{ >{\raggedright\arraybackslash}p{6cm} ccc ccc ccc }
    \toprule & 
    \multicolumn{3}{c}{\textbf{F1-score}} & \multicolumn{3}{c}{\textbf{Precision}} & \multicolumn{3}{c}{\textbf{Recall}} \\
    \cmidrule(lr){2-4} \cmidrule(lr){5-7} \cmidrule(lr){8-10}
    \textbf{Transaction method} & \textbf{Graph.} & \textbf{Events} & \textbf{Comb.} & \textbf{Graph.} & \textbf{Events} & \textbf{Comb.} & \textbf{Graph.} & \textbf{Events} & \textbf{Comb.} \\ 
    \midrule 
Transfer & 0.976 & 0.969 & \textbf{0.982} & 0.959 & 0.945 & \textbf{0.971} & \textbf{0.994} & \textbf{0.994} & 0.993 \\
Execute & 0.964 & 0.927 & \textbf{0.969} & \textbf{0.964} & 0.892 & 0.962 & 0.964 & 0.965 & \textbf{0.977} \\
Swap & 0.986 & 0.855 & \textbf{0.988} & \textbf{0.987} & 0.942 & 0.982 & 0.985 & 0.783 & \textbf{0.995} \\
Swap Exact ETH For Tokens Supporting Fee On Transfer Tokens & 0.627 & 0.799 & \textbf{0.855} & 0.485 & 0.727 & \textbf{0.819} & 0.888 & 0.886 & \textbf{0.894} \\
Swap Exact Tokens For ETH Supporting Fee On Transfer Tokens & 0.604 & 0.929 & \textbf{0.983} & 0.522 & 0.955 & \textbf{0.984} & 0.717 & 0.905 & \textbf{0.983} \\
0x0162e2d0 & 0.467 & \textbf{1.000} & \textbf{1.000} & 0.915 & \textbf{1.000} & \textbf{1.000} & 0.313 & \textbf{1.000} & \textbf{1.000} \\
Deposit & 0.735 & 0.895 & \textbf{0.917} & 0.961 & \textbf{0.990} & 0.960 & 0.595 & 0.817 & \textbf{0.878} \\
0x75713a08 & 0.393 & \textbf{1.000} & \textbf{1.000} & 0.396 & \textbf{1.000} & \textbf{1.000} & 0.390 & \textbf{1.000} & \textbf{1.000} \\
Transfer From & 0.144 & 0.228 & \textbf{0.298} & \textbf{0.993} & 0.630 & 0.928 & 0.078 & 0.139 & \textbf{0.178} \\
0x415565b0 & 0.990 & \textbf{0.999} & \textbf{0.999} & 0.985 & \textbf{0.999} & \textbf{0.999} & 0.996 & \textbf{1.000} & \textbf{1.000} \\
Unxswap By Order Id & 0.307 & \textbf{1.000} & \textbf{1.000} & 0.578 & 0.999 & \textbf{1.000} & 0.209 & \textbf{1.000} & 0.999 \\
Withdraw & 0.805 & 0.858 & \textbf{0.942} & 0.858 & \textbf{0.996} & 0.976 & 0.758 & 0.754 & \textbf{0.910} \\
Claim & 0.794 & 0.805 & \textbf{0.892} & 0.777 & \textbf{0.984} & 0.855 & 0.812 & 0.681 & \textbf{0.931} \\
Propose Block & \textbf{1.000} & \textbf{1.000} & \textbf{1.000} & \textbf{1.000} & \textbf{1.000} & \textbf{1.000} & \textbf{1.000} & \textbf{1.000} & \textbf{1.000} \\
Execute F Fs Yo & 0.911 & 0.237 & \textbf{0.920} & \textbf{0.998} & 0.682 & \textbf{0.998} & 0.837 & 0.143 & \textbf{0.854} \\
Multicall & 0.702 & 0.840 & \textbf{0.882} & 0.840 & \textbf{0.937} & 0.908 & 0.602 & 0.762 & \textbf{0.858} \\
Mint & 0.936 & 0.318 & \textbf{0.951} & 0.915 & 0.886 & \textbf{0.922} & 0.958 & 0.194 & \textbf{0.982} \\
Deposit ETH & 0.652 & \textbf{0.989} & 0.903 & 0.489 & \textbf{0.994} & 0.824 & 0.981 & 0.983 & \textbf{0.999} \\
Stake & 0.478 & 0.779 & \textbf{0.941} & 0.429 & \textbf{0.978} & 0.938 & 0.539 & 0.647 & \textbf{0.945} \\
Sigma Buy & 0.433 & \textbf{1.000} & \textbf{1.000} & 0.451 & \textbf{1.000} & \textbf{1.000} & 0.417 & \textbf{1.000} & \textbf{1.000} \\
Swap Exact ETH For Tokens & 0.155 & 0.000 & \textbf{0.337} & 0.659 & 0.000 & \textbf{0.915} & 0.088 & 0.000 & \textbf{0.206} \\
Bond Withdrawal & 0.975 & \textbf{1.000} & \textbf{1.000} & 0.951 & \textbf{1.000} & \textbf{1.000} & \textbf{1.000} & \textbf{1.000} & \textbf{1.000} \\
Sigma Sell & 0.626 & \textbf{1.000} & \textbf{1.000} & 0.934 & \textbf{1.000} & \textbf{1.000} & 0.471 & \textbf{1.000} & \textbf{1.000} \\
Swap Exact Tokens For Tokens & 0.896 & 0.794 & \textbf{0.954} & 0.879 & 0.677 & \textbf{0.921} & 0.913 & 0.961 & \textbf{0.990} \\
0x00000195 & 0.733 & \textbf{1.000} & \textbf{1.000} & 0.579 & \textbf{1.000} & \textbf{1.000} & \textbf{1.000} & \textbf{1.000} & \textbf{1.000} \\
Send Message & 0.288 & \textbf{1.000} & \textbf{1.000} & \textbf{1.000} & \textbf{1.000} & \textbf{1.000} & 0.168 & \textbf{1.000} & \textbf{1.000} \\
0xa0000000 & 0.840 & 0.563 & \textbf{0.861} & 0.738 & 0.397 & \textbf{0.761} & 0.975 & 0.967 & \textbf{0.990} \\
Uniswap V3Swap To & 0.407 & 0.970 & \textbf{0.986} & 0.500 & \textbf{0.975} & \textbf{0.975} & 0.344 & 0.965 & \textbf{0.998} \\
Buy With Eth & 0.984 & 0.991 & \textbf{0.992} & 0.999 & \textbf{1.000} & \textbf{1.000} & 0.969 & 0.983 & \textbf{0.984} \\
0x263fd4a1 & 0.985 & 0.000 & \textbf{0.991} & 0.971 & 0.000 & \textbf{0.982} & \textbf{1.000} & 0.000 & \textbf{1.000} \\
    \bottomrule 
    \end{tabular}} 
\end{table}

\subsection{Method-Level Performance}
\label{app:method-performance}

To better understand why the combined representation performs best, Table~\ref{tab:app_method-performance} compares the per-method F1-scores, precision, and recall obtained using the graphlet-based \textit{all features}, the \textit{top 1,000 events}, and their \textit{combined} models. The relative effectiveness of the two individual representations varies considerably across transaction methods. 
%Of the 30 methods, graphlet-based features achieve the highest F1-score for 12 methods, event-log features perform better for 17 methods, and both representations achieve the same score for one method.
Graphlet features alone achieve a higher F1-score than event log features for 9 of the 30 methods, event log features for 20 methods. Both reach the same F1-score of 1.000 for \texttt{Propose Block}.

Graphlet-based features perform particularly well for methods whose functionality produces distinctive token-transfer structures. For example, graphlets outperform event log counts for \texttt{Swap}, achieving F1-scores of 0.986 and 0.855, respectively. Similar patterns are observed for \texttt{Execute F Fs Yo} (0.911 versus 0.237), \texttt{Mint} (0.936 versus 0.318), and \texttt{0xa0000000} (0.840 versus 0.563). Event log features alone fail to identify \texttt{0x263fd4a1} and \texttt{Swap Exact ETH For Tokens} (F1-score of 0.000), whereas graphlet features reach 0.985 and 0.155, respectively.

In contrast, event-log features perform substantially better for methods with distinctive emitted event signatures. For example, the event log features achieve an F1-score of 0.970 for \texttt{Uniswap V3Swap To}, compared with 0.407 for graphlet-based representation. Event logs also outperform graphlets for \texttt{Swap Exact Tok-} \texttt{ens For ETH Supporting Fee On Transfer Tokens} (0.929 versus 0.604), \texttt{De-} \texttt{posit ETH} (0.989 versus 0.652), and \texttt{Multicall} (0.840 versus 0.702). They also outperform graphlets for \texttt{Withdraw} (0.858 versus 0.805) and \texttt{Claim} (0.805 versus 0.794), and several unlabeled and protocol-specific methods (\texttt{0x0162e2d0}, \texttt{0x75713a08}, \texttt{Unxswap By Order Id}, \texttt{Sigma Buy}, \texttt{Sigma Sell}, \texttt{Send Message}, and \texttt{0x00000195}) are identified perfectly by event logs (F1-score of 1.000), where-as graphlets alone reach only 0.288 to 0.733.

The combined representation achieves a higher F1-score than either individual representation for 19 of the 30 transaction methods and matches the best individual score for 10 additional methods. Particularly large improvements are observed for \texttt{Stake}, for which the combined F1-score reaches 0.941, compared with 0.478 for graphlets and 0.779 for event logs. The combined representation also improves performance for \texttt{Deposit} (0.917 versus a best individual score of 0.895), \texttt{Withdraw} (0.942 versus 0.858), and \texttt{Claim} (0.892 versus 0.805). The only method for which it falls below the best individual representation is \texttt{Deposit ETH} (0.903 versus 0.989 for event logs), where precision drops from 0.994 to 0.824 while recall slightly increases from 0.983 to 0.999.

%The main exception is \texttt{Deposit ETH}, for which the event log representation achieves an F1-score of 0.979, compared with 0.904 for the combined representation. Although combining the features increases recall from 0.962 to 1.000, precision decreases from 0.997 to 0.825. This suggests that the event signature already identifies this method very reliably and that adding graphlet features introduces ambiguity with structurally similar transactions.

The method-level results confirm that graphlet-based and event log features capture complementary aspects of transaction behavior. Their combination improves performance, particularly for methods that cannot be identified reliably using either transaction topology or emitted events alone.

\section{Interpretable Machine Learning}
\label{app:shap}
\subsection{Feature-level SHAP analysis}
\label{app:shap-features}
We first examine the relative contribution of graphlet-based and event-log features to the classifier's predictions. For each of the top 15 methods, Table~\ref{tab:app_top_features_by_method} lists the top 10 features with the largest mean absolute SHAP contributions over the corresponding explainability set $\mathcal{T}_m$. Note that $\mathcal{T}_m$ size depends on the related method support in the test set.
For each method, we compare the cumulative absolute SHAP contributions of graphlet-based and event-log features and report the graphlet participation ratio (GPR), the ratio of cumulative absolute SHAP values of graphlet-based features to the cumulative absolute SHAP values of all features. We also consider graphlet dominance (\% graphlet), defined as the proportion of transactions in $\mathcal{T}_m$ that exhibit higher cumulative absolute SHAP values for graphlet features than for event log features.

Graphlet-based features dominate the explanations for most methods with GPR above $50\%$. Highest ratios are attributed to \texttt{Mint} and \texttt{Transfer from} ($80\%$), followed by swap-related methods \texttt{Swap Exact ETH For Tokens Suppor-} \texttt{ting
Fee On Transfer Tokens} ($75\%$) and \texttt{Swap Exact Tokens for ETH Sup-} \texttt{porting
Fee On Transfer Tokens} ($69\%$) and \texttt{Transfer} ($67.2\%$). Then \texttt{Exec-} \texttt{ute}, \texttt{Deposit}, \texttt{Withdraw}, \texttt{Multicall} and \texttt{Swap} are around $60\%$, and \texttt{Claim} reaches $53.8\%$. These results indicate that, for most considered methods, the structural information encoded by graphlet-based features contributes more stron-gly to the model predictions than the event-log features alone.
An important exception is \texttt{Unxswap By Order Id}, with a GPR of only $29.4\%$. Its predictions, therefore, rely primarily on event log features, possibly because method-specific parameters such as order identifiers and routing information are not captured by local transfer topology. Three unlabeled methods \texttt{0x415565b0}, \texttt{0x0162e2d0} and \texttt{0x75713a08} are associated with low GPR ($\in [33,44]$).

Regarding graphlet dominance scores (\% graphlet), most of the considered methods show a strong graphlet dominance, with ratios exceeding $90\%$ in many cases. Methods associated with Swap operations (\texttt{Swap}, \texttt{Swap Exact ETH For Tokens Supporting Fee On Transfer Tokens}, and \texttt{Swap Exact Tokens For ETH Supporting Fee On Transfer Tokens}) show graphlet dominance for $66\%$, $100\%$, and $99.9\%$ of the transactions in their respective $\mathcal{T}_m$. \texttt{Mint} and \texttt{Execute} also show strong graphlet dominance, with ratios of $\approx 97\%$, while \texttt{Claim} and \texttt{Multicall} show intermediate ratios ranging from $72\%$ to $76\%$. \texttt{Deposit} presents a graphlet dominance of $84\%$. In contrast, \texttt{Unxswap By Order Id} shows a very low contribution of graphlet features, with graphlet dominance observed for only $0.1\%$ of the transactions, which indicates that this method is identified almost entirely from its event logs. This behavior may be related to the highly structured nature of this operation, which explicitly relies on an order identifier and dedicated swap-routing parameters. Finally, the unlabeled methods are heterogeneous: \texttt{0x0162e2d0} and \texttt{0x75713a08} show a very low graphlet dominance ($\approx 4\%$) and \texttt{0x415565b0} a low one ($36\%$), whereas \texttt{0xa0000000}, \texttt{0x263fd4a1}, and \texttt{0x00000195} exceed $93\%$.

Features are represented by a variety of feature families in the top 10 features. Top graphlet-based features are governed by local configurations of addresses and contracts: \emph{*\_Address} and \emph{*\_Contract}. We also observe information on the associated token type, such as \emph{0\_1\_ETH, \emph{0\_1\_Cryptocurrency}}, especially in the context of swap-related methods, \texttt{Withdraw}, \texttt{Claim}, \texttt{Transfer}, and \texttt{Transfer from}. Specific graphlet signatures also provide interpretable examples of how local topology can characterize DeFi methods. \texttt{Mint} method shows a strong contribution from \emph{0\_Null} graphlet feature, which is consistent with the creation of a new token. The presence of the \emph{0\_1\_erc721} highlights the method relation in the NFT ecosystem. In contrast, \texttt{Unxswap By Order Id} is characterized by the \emph{0\_1\_Protocol} feature, reflecting a router- or protocol-level structure.

Table~\ref{tab:app_top_features_by_method_ext} extends the previous analysis to 15 additional top DeFi methods and unlabeled function signatures. The results confirm the predominance of graphlet-based information for several operations. In particular, \texttt{Execute F Fs Yo}, \texttt{0xa0000000}, \texttt{0x263fd4a1}, \texttt{Swap Exact Tokens For Tokens}, and \texttt{Swap Exact ETH For Tokens} all show high GPR values, ranging from $72\%$ to almost $79\%$. These methods also exhibit high graphlet dominance, with more than $93\%$ of the transactions showing a larger cumulative absolute contribution from graphlet-based than event-log features. This further supports the relevance of local transaction topology for characterizing specific DeFi operations.

The additional swap-related methods also illustrate heterogeneity in the relative contribution of graphlet-based features. \texttt{Swap Exact Tokens For Tokens} and \texttt{Swap Exact ETH For Tokens} show strong graphlet contributions, with GPR values of $73\%$ and $76\%$, respectively, and graphlet dominance above $99\%$. In contrast, \texttt{Uniswap V3Swap To} exhibits a much weaker graphlet contribution, with a GPR of only $35\%$ and graphlet dominance in $4.3\%$ of transactions.

Some methods show a more balanced contribution between the two feature families. \texttt{Send Message}, for example, has a GPR of $50.8\%$ and graphlet dominance for $55.2\%$ of transactions. Similarly, \texttt{Propose Block} presents a GPR close to $55\%$. \texttt{Stake} is another notable case, with a GPR of $43.7\%$. Conversely, \texttt{Buy With Eth}, \texttt{Sigma Buy}, and \texttt{Sigma Sell} show weak graphlet contributions, with GPR values below $40\%$. These cases illustrate that local topology does not provide a universal explanation for all DeFi methods and that some operations are more directly identified through their event log or contract-specific information. The unlabeled methods provide additional evidence that graphlet-based features can reveal structural distinctions even when explicit method information is unavailable. The method \texttt{0xa0000000} shows a strong graphlet contribution, with a GPR of $73.1\%$, while \texttt{0x263fd4a1} reaches $78.9\%$. The latter is particularly characterized by \emph{3901\_3902\_ETH}, \emph{0\_1\_ETH}, and several address and contract configurations, suggesting a distinctive topology involving ETH transfers. In contrast, \texttt{0x00000195} is associated with a GPR of $58.6\%$. These results reinforce the observation that unlabeled methods can exhibit distinct and interpretable structural signatures despite the absence of explicit semantic labels.

Finally, the composition of the top graphlet features provides additional insight into the type of structural information exploited by the classifier. Address- and contract-based configurations remain frequent across methods, while token-type features such as \emph{0\_1\_ETH}, \emph{0\_1\_ETH/WETH}, and \emph{0\_1\_Cryptocurrency} appear particularly often in exchange, deposit, and withdrawal operations.

\subsection{Graphlet signatures}
\label{app:signatures}
\begin{figure}
    \centering
        \includegraphics[width=\linewidth]{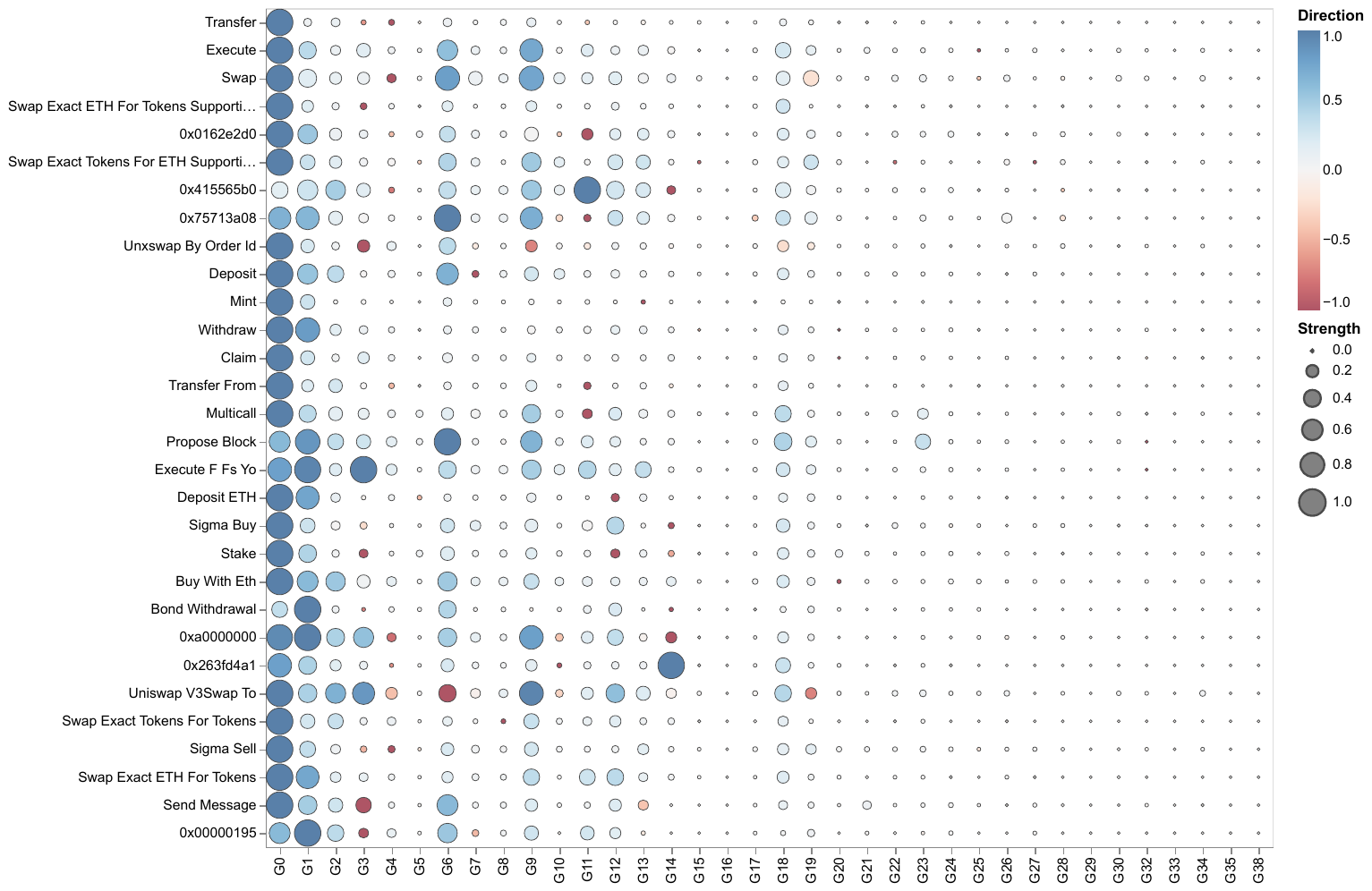}
    \caption{blue}{Method-specific graphlet signatures for all 30 methods (see Fig.~\ref{fig:heatmap} for the notation). $G31$, $G36$, $G37$, and $G39$ are absent from our data.}
    \label{fig:app_heatmap}
\end{figure}

In the following analysis, we consider the method heatmap of aggregated graphlet importance. While Fig.~\ref{fig:app_heatmap} presents, for each method, the distribution of directional and signed importance ($\tilde{\Pi}^D$), it also provides insight into the magnitude of this score ($\tilde{\Pi}^S$). A positive direction indicates a positive effect on the model's prediction of the considered method relative to the baseline. In contrast, interpreting negative scores as counter-signatures would require a specific methodology to distinguish the absence of a structural pattern from the presence of an alternative structural signature.

$G0$ makes a strong contribution to the prediction of most methods, with some notable exceptions, such as unlabeled methods and \texttt{Bond Withdrawal}. The strong contribution of $G0$ for \texttt{Transfer} is particularly consistent with the dyadic nature of direct token transfers.

Swap-related methods exhibit substantial positive graphlet contributions across $G0$–$G3$, consistent with the presence of intermediate components involved in token exchange operations. Similar contributions from small graphlets are also observed for \texttt{Deposit}, \texttt{Withdraw}, \texttt{Stake}, and \texttt{Execute}. However, some methods exhibit signatures extending beyond these smallest graphlets. In particular, \texttt{Uniswap V3Swap To} shows strong contributions from $G9$, $G3$, $G2$, and $G12$, while \texttt{Swap} presents particularly strong contributions from $G6$ and $G9$. This suggests that graphlet signatures can distinguish different types of exchange operations despite their common semantic function.

Larger graphlets also contribute to the signatures of specific methods, notably $G6$, $G9$, $G12$, $G13$, $G14$, $G18$, and $G19$. These contributions are particularly visible for \texttt{Multicall}, \texttt{Execute}, \texttt{Stake}, and several swap variants, suggesting that some DeFi operations are characterized by more extended local topological structures rather than only small-scale motifs. In contrast, methods such as \texttt{Mint} and \texttt{Claim} exhibit comparatively simpler signatures, dominated by $G0$ and $G1$.

Unlabeled methods exhibit distinct graphlet signatures from one another despite the absence of explicit method information. For example, \texttt{0x415565b0} is characterized by strong contributions from $G11$, $G9$, and $G6$, whereas \texttt{0x75713a08} shows stronger contributions from $G6$ and $G9$. The method \texttt{0xa0000000} presents contributions from $G9$, $G6$, and $G3$, while \texttt{0x263fd4a1} is particularly characterized by $G14$. Finally, \texttt{0x00000195} shows contributions from $G6$, $G9$, $G11$, and $G12$. These distinct signatures suggest that graphlet-based features can differentiate otherwise unlabeled methods according to their local transaction topology.

\begin{table}[t]
\centering
\caption{Top 10 most important features per method by mean absolute SHAP value (graphlet-based features in bold, importance values in parentheses), with the explainability set size $|\mathcal{T}_{\mathrm{m}}|$, the graphlet dominance (\% graphlet), and the graphlet participation ratio (GPR, \%). Part 1 of 2: first 15 of the 30 methods.}
\label{tab:app_top_features_by_method}
\scriptsize	
\begin{tabularx}{\textwidth}{p{1.5cm} p{0.8cm} p{0.8cm} p{1.1cm} X}
\toprule
\textbf{Method} & $|\mathcal{T}_{\mathrm{m}}|$ & \textbf{\% graphlet} & \textbf{GPR} & \textbf{Top 10 features (importance)} \\
\midrule
Transfer & 586 & 99.6 & 67.2 & \textbf{0\_Contract (0.25)}, Approval (0.09), Swap\_4 (0.08), Sync (0.08), Swap\_5 (0.06), \textbf{0\_1\_ETH/WETH (0.03)}, \textbf{1\_Contract (0.03)}, Deposit\_7 (0.03), Permit (0.02), \textbf{14\_Address (0.02)} \\
Execute & 1000 & 96.8 & 62.28 & Permit (0.11), Swap\_3 (0.08), \textbf{26\_Address (0.07)}, \textbf{1\_Contract (0.05)}, TransformedERC20 (0.05), OrderRecord (0.04), BridgeFill (0.03), \textbf{0\_Contract (0.02)}, \textbf{13\_Address (0.02)}, \textbf{16\_Address (0.02)} \\
Swap & 1000 & 66.4 & 58.65 & Swap\_3 (0.35), \textbf{26\_Address (0.07)}, Permit (0.05), \textbf{0\_Contract (0.05)}, TransformedERC20 (0.03), \textbf{1\_Contract (0.03)}, \textbf{16\_Address (0.03)}, \textbf{13\_Address (0.02)}, ClientData (0.02), 0 (0.02) \\
Swap Exact ETH For Tokens ... & 26 & 100 & 74.99 & \textbf{1\_Contract (0.09)}, \textbf{0\_Contract (0.07)}, \textbf{0\_1\_Cryptocurrency (0.06)}, Deposit\_7 (0.06), \textbf{0\_1\_ETH (0.05)}, Permit (0.03), \textbf{28\_Address (0.02)}, Swap\_4 (0.02), \textbf{51\_52\_erc20 (0.02)}, \textbf{2\_Address (0.02)} \\
0x0162e2d0 & 49 & 4.3 & 41.64 & 0x72...499f (0.44), 0x9f...c420 (0.11), \textbf{1\_Contract (0.06)}, Swap\_3 (0.03), \textbf{4\_Address (0.02)}, Deposit\_7 (0.02), OrderRecord (0.02), \textbf{0\_Contract (0.02)}, Withdrawal (0.02), \textbf{26\_Address (0.02)} \\
Swap Exact Tokens For ETH ... & 250 & 99.9 & 68.61 & Swap\_4 (0.06), Permit (0.05), Swap\_5 (0.05), \textbf{0\_1\_ETH (0.05)}, \textbf{0\_Contract (0.05)}, \textbf{26\_Address (0.04)}, \textbf{1\_Contract (0.04)}, TransformedERC20 (0.04), \textbf{53\_Address (0.03)}, Withdrawal (0.03) \\
0x415565b0 & 1000 & 36.1 & 43.99 & TransformedERC20 (0.42), BridgeFill (0.15), \textbf{31\_Address (0.05)}, Swap\_3 (0.04), OtcOrderFilled (0.02), \textbf{4\_Address (0.02)}, \textbf{35\_Address (0.02)}, \textbf{5\_Address (0.02)}, \textbf{26\_Address (0.02)}, \textbf{7\_Address (0.02)} \\
0x75713a08 & 121 & 4.22 & 33.47 & Sell (0.36), 0x72...499f (0.17), Withdrawal (0.06), Swap\_3 (0.03), \textbf{26\_Address (0.02)}, Permit (0.02), BridgeFill (0.02), \textbf{13\_Contract (0.02)}, \textbf{3\_Address (0.02)}, Swap\_4 (0.01) \\
Unxswap By Order Id & 369 & 0.1 & 29.44 & OrderRecord (0.63), Swap\_4 (0.08), Sync (0.03), Permit (0.03), \textbf{0\_Contract (0.03)}, Swap\_5 (0.02), \textbf{0\_1\_Protocol (0.02)}, \textbf{1\_Contract (0.01)}, Swap\_3 (0.01), TransformedERC20 (0.01) \\
Deposit & 294 & 84 & 62.44 & \textbf{0\_Contract (0.11)}, Deposit\_7 (0.10), \textbf{1\_Contract (0.06)}, Swap\_5 (0.05), \textbf{4\_Address (0.05)}, Deposit\_5 (0.04), LogDeposit (0.04), \textbf{14\_Address (0.03)}, Swap\_4 (0.03), Permit (0.03) \\
Mint & 174 & 97.3 & 79.07 & \textbf{0\_Null (0.23)}, \textbf{1\_Contract (0.11)}, \textbf{4\_Address (0.08)}, \textbf{0\_Contract (0.05)}, \textbf{0 (0.05)}, Swap\_3 (0.03), Sync (0.03), Swap\_5 (0.03), Swap\_4 (0.02), \textbf{0\_1\_erc721 (0.02)} \\
Withdraw & 293 & 69.9 & 61.22 & \textbf{1\_Contract (0.08)}, \textbf{4\_Address (0.07)}, \textbf{0\_1\_Cryptocurrency (0.06)}, \textbf{2\_Address (0.05)}, Withdrawn\_3 (0.04), Swap\_5 (0.03), Withdtraw (0.03), WithdrawFromLockupStream (0.03), Withdrawal (0.03), Withdraw\_5 (0.03) \\
Claim & 174 & 71.8 & 53.83 & \textbf{1\_Contract (0.11)}, TokenSent (0.05), 0x9d...a305 (0.04), Claimed\_6 (0.04), 0x46...911f (0.03), Swap\_5 (0.03), 0xa3...7852 (0.03), \textbf{0\_1\_Cryptocurrency (0.03)}, Claimed\_9 (0.03), \textbf{4\_Address (0.03)} \\
Transfer From & 20 & 100 & 79.01 & \textbf{0\_1\_ETH/WETH (0.11)}, \textbf{0\_Contract (0.09)}, \textbf{0\_1\_erc721 (0.07)}, Swap\_5 (0.06), \textbf{5\_Address (0.06)}, Approval (0.04), \textbf{1\_Contract (0.04)}, \textbf{2\_Address (0.04)}, Swap\_4 (0.03), Transfer\_3 (0.03) \\
Multicall & 450 & 76.2 & 63.31 & \textbf{1\_Contract (0.08)}, Collect (0.07), Burn (0.07), Swap\_4 (0.05), IncreaseLiquidity (0.04), Swap\_3 (0.04), \textbf{0\_Contract (0.04)}, \textbf{26\_Address (0.03)}, Swap\_5 (0.03), Deposit\_7 (0.02) \\
\bottomrule
\end{tabularx}
\end{table}

\begin{table}[t]
\centering
\caption{Same as Table~\ref{tab:app_top_features_by_method}. Part 2 of 2: remaining 15 of the 30 methods.}
\label{tab:app_top_features_by_method_ext}
\scriptsize	
\begin{tabularx}{\textwidth}{p{1.5cm} p{0.8cm} p{0.8cm} p{1.1cm} X}
\toprule
\textbf{Method} & $|\mathcal{T}_{\mathrm{m}}|$ & \textbf{\% graphlet} & \textbf{GPR} & \textbf{Top 10 features (importance)} \\
\midrule
Propose Block & 9 & 64.2 & 54.93 & 0x98...f0b1 (0.26), Swap\_3 (0.05), Permit (0.04), \textbf{4\_Address (0.03)}, \textbf{16\_Address (0.03)}, \textbf{69\_Contract (0.02)}, \textbf{0\_1\_Cryptocurrency (0.02)}, Swap\_4 (0.02), OrderRecord (0.02), TransformedERC20 (0.02) \\
Execute F Fs Yo & 139 & 95.1 & 72.34 & \textbf{7\_Address (0.08)}, Withdrawal (0.04), 32 (0.03), Swap\_3 (0.03), TokenExchange (0.03), Swap\_5 (0.03), \textbf{0\_1\_ETH (0.03)}, \textbf{0\_Contract (0.03)}, \textbf{2\_Address (0.03)}, Sync (0.02) \\
Deposit ETH & 22 & 89.9 & 61.89 & Deposit\_3 (0.13), \textbf{4\_Address (0.10)}, \textbf{0\_Contract (0.09)}, \textbf{0\_1\_ETH (0.05)}, Swap\_4 (0.05), ETHBridgeInitiated (0.04), Permit (0.04), \textbf{0 (0.03)}, TransformedERC20 (0.03), Approval (0.02) \\
Sigma Buy & 50 & 38.4 & 39.59 & 0x24...4e08 (0.43), 0x33...feb3 (0.16), \textbf{1\_Contract (0.05)}, Swap\_3 (0.03), \textbf{33\_Address (0.03)}, TransformedERC20 (0.02), \textbf{0\_1\_ETH/WETH (0.02)}, \textbf{0\_1\_ETH (0.02)}, \textbf{0\_Contract (0.02)}, \textbf{25\_Address (0.02)} \\
Stake & 105 & 41 & 43.74 & Staked (0.11), 0x22...c42e (0.08), Stake\_2 (0.08), Stake\_4 (0.07), Staked\_2 (0.05), selfStake (0.04), Swap\_3 (0.04), \textbf{0\_Contract (0.04)}, 0x7d...7692 (0.03), \textbf{0\_1\_Cryptocurrency (0.03)} \\
Buy With Eth & 10 & 0 & 33.89 & TokensBoughtAndStaked (0.32), TokensBought (0.23), Swap\_3 (0.04), \textbf{0\_1\_ETH (0.03)}, \textbf{1\_Contract (0.02)}, \textbf{5\_Address (0.02)}, TransformedERC20 (0.02), Permit (0.02), OrderRecord (0.02), \textbf{0 (0.02)} \\
Bond Withdrawal & 10 & 97 & 51.72 & WithdrawalBonded (0.41), \textbf{2\_Address (0.07)}, \textbf{0\_1\_ETH (0.04)}, 0 (0.04), TransformedERC20 (0.03), \textbf{3\_4\_ETH (0.03)}, \textbf{4\_Contract (0.03)}, \textbf{16\_Address (0.03)}, \textbf{33\_Address (0.02)}, \textbf{4\_Address (0.02)} \\
0xa0000000 & 210 & 93.4 & 73.11 & Swap\_4 (0.07), OrderFilled (0.06), \textbf{4\_Address (0.05)}, \textbf{7\_Address (0.05)}, \textbf{2\_Address (0.04)}, \textbf{26\_Address (0.04)}, Swap\_3 (0.03), \textbf{0\_1\_Cryptocurrency (0.03)}, \textbf{16\_Address (0.02)}, \textbf{1\_Contract (0.02)} \\
0x263fd4a1 & 18 & 100 & 78.87 & \textbf{3901\_3902\_ETH (0.11)}, \textbf{39\_Contract (0.07)}, \textbf{0\_1\_ETH (0.07)}, Deposit\_7 (0.06), \textbf{2\_Address (0.03)}, \textbf{13\_Address (0.03)}, \textbf{1\_Contract (0.03)}, \textbf{26\_Address (0.03)}, Swap\_4 (0.03), \textbf{0\_1\_Cryptocurrency (0.03)} \\
Uniswap V3Swap To & 155 & 4.3 & 34.95 & OrderRecord (0.50), Swap\_4 (0.09), Swap\_5 (0.09), Sync (0.05), \textbf{0\_Contract (0.02)}, \textbf{26\_Address (0.02)}, \textbf{7\_Address (0.02)}, \textbf{0\_1\_Protocol (0.02)}, \textbf{1\_Contract (0.01)}, TransformedERC20 (0.01) \\
Swap Exact Tokens For Tokens & 122 & 99.6 & 72.96 & \textbf{1\_Contract (0.11)}, \textbf{0\_Contract (0.09)}, Swap\_5 (0.06), Sync (0.04), Swap\_4 (0.04), Permit (0.04), \textbf{13\_Address (0.04)}, \textbf{26\_Address (0.03)}, \textbf{4\_Address (0.03)}, \textbf{5\_Address (0.03)} \\
Sigma Sell & 88 & 8.5 & 33.78 & 0x24...4e08 (0.35), 0xbb...db16 (0.23), \textbf{0\_Contract (0.04)}, Swap\_4 (0.03), \textbf{0\_1\_Cryptocurrency (0.03)}, Swap\_3 (0.03), TransformedERC20 (0.02), OrderRecord (0.01), Permit (0.01), 0x65...5b72 (0.01) \\
Swap Exact ETH For Tokens & 31 & 100 & 76 & \textbf{1\_Contract (0.11)}, \textbf{3\_4\_ETH (0.06)}, Deposit\_7 (0.05), \textbf{32 (0.05)}, \textbf{33\_Address (0.04)}, \textbf{2\_Address (0.04)}, Swap\_3 (0.04), Swap\_4 (0.03), \textbf{0\_Contract (0.02)}, \textbf{0\_1\_ETH (0.02)} \\
Send Message & 38 & 55.2 & 50.82 & MessageSent\_2 (0.19), MessageSent (0.13), SentMessage (0.12), \textbf{0 (0.06)}, Permit (0.04), \textbf{1\_Contract (0.04)}, \textbf{16\_Address (0.03)}, \textbf{0\_Contract (0.03)}, \textbf{7\_Address (0.02)}, \textbf{6 (0.02)} \\
0x00000195 & 5 & 99.9 & 58.6 & SocketBridge (0.19), TransactionDeposited (0.14), \textbf{0 (0.05)}, \textbf{0\_Contract (0.05)}, Swap\_3 (0.05), \textbf{2\_Address (0.04)}, \textbf{13\_Address (0.03)}, \textbf{32 (0.02)}, \textbf{4\_Contract (0.02)}, \textbf{4\_Address (0.02)} \\
\bottomrule
\end{tabularx}
\end{table}

\end{document}